\documentclass[10pt,twocolumn,showpacs  ,secnumarabic, nobibnotes,  nofootinbib, aps,
prd]{revtex4}%
\usepackage{amsmath}
\usepackage{amssymb}
\usepackage{bm}
\usepackage{graphicx}
\usepackage{caption}
\usepackage{amsfonts}%
\providecommand{\U}[1]{\protect\rule{.1in}{.1in}}
\def\EMPH#1{{#1}}
\begin{document}
\title{A statistical analysis of two-dimensional patterns and its application to astrometry}
\author{Petr Zavada and Karel P\'{\i}\v{s}ka}
\affiliation{Institute of Physics AS CR, Na Slovance 2, CZ-182 21 Prague 8, Czech Republic}
\email{zavada@fzu.cz}

\begin{abstract}
{Here we develop a general statistical procedure for the analysis of finite two-dimensional (2D) patterns inspired
by the analysis of heavy-ion data. The method is used in the study
of publicly available data obtained by the Gaia-ESA mission. We prove that the
procedure can be sensitive to the limits of accuracy of measurement, and
can also clearly identify the real physical effects on the large background of
random distributions. As an example, the method confirms the presence of binary
and ternary star systems in the studied data. At the same time, the possibility
of the statistical detection of the gravitational microlensing effect is discussed.}
\end{abstract}

\keywords{Methods: statistical -- Methods: data analysis -- Surveys -- Astrometry --
Stars: binaries -- Gravitational lensing: micro}
%\pacs{98.80.-d, 97.10.Yp, 95.12.Fe}

\maketitle

\section{Introduction}

The motivation of the present study was to modify and generalize the known
method for analyzing anisotropic flow in relativistic nuclear collisions
\cite{Voloshin:1994mz,Poskanzer:1998yz}, which has been effectively
\EMPH{applied in many studies, for example, recently in \cite{Adam:2016izf}.}
The method
is based on the use of the Fourier expansion of azimuthal distributions of
produced particles and allows us to obtain important information on the mechanism
of nuclear collisions. However, mathematical formalism of this method is more
general and can be used after minor modifications even for quite different
kinds of analysis. Our present idea is focused primarily on astrometry.
Recently, some similarity between spiral structures in galactic patterns and
heavy ion collisions has been discussed \cite{Rustamov:2016xtv}. However,
our approach is different. We make use of the formalism of the Fourier
analysis of nuclear collisions simply as a tool and we do not consider a
common physics that could bridge the two different fields. Moreover, Fourier
analysis is only one of two methods that we work with.

In our approach the astrometric data are decomposed into a set of limited,
finite star patterns whose parameters are statistically analyzed. These
parameters define characteristics of the patterns and represent statistical
deviations from the uniform distribution of stars; for example, a tendency to
display scale-dependent clustering or anti-clustering. The aim is to find and
interpret these deviations. The input data are taken from the Gaia DR1 catalog
\cite{gaia1,gaia2}.

In Sect. \ref{soa} we define some useful terms and explain the essence of our
task in more detail. The general description of the Fourier analysis modified
for application to astrometry is presented in Sect. \ref{aac}. Then in
Sect. \ref{dod} we describe a complementary statistical method for analysis of
the patterns of stars, which deals with angular distances and is important for
identification of binary and ternary star systems. The results obtained from the
application of both methods to the Gaia data are presented and discussed
in Sect. \ref{ata}. A discussion on the gravitational microlensing effect and
the conditions of its observation is given Sect. \ref{wdn}. A brief summary of the
paper is presented in Sect. \ref{sum}.

\section{Methods}

\label{meth}

\subsection{Subject of analysis}

\label{soa}Let us consider a region of the sphere of the galactic reference
frame. We  study the patterns of the stars inside the circles of the same
angular radius as in Fig.\ref{faa1}a that cover the chosen region. The
circular shape is essential for application of our methods and the stars among
the circles are not used for statistical analysis. First, in accordance
with the figure, we define one \textit{event} of the \textit{multiplicity} $M$
as a set of stars with angular positions $\left\{  l_{i},b_{i}\right\}  ,\quad
i=1,...M$ \ inside one circle with the center $(l_{0},b_{0})$ and a small
angular radius $\rho$ (introduced \textit{terms} are inspired by particle
physics). The letter $l (b)$ represents the galactic longitude (latitude). The
positions inside the circle (event) can be represented equivalently by the three-dimensional (3D)
unit vectors $\mathbf{n}_{\alpha}~$as sketched in Fig.\ref{faa1}b:
\begin{equation}
\mathbf{n}_{\alpha}=(\cos b_{\alpha}\cos l_{\alpha},\cos b_{\alpha}\sin
l_{\alpha},\sin b_{\alpha}),\qquad\alpha=0,...M. \label{sa1}%
\end{equation}
Subsequently, the event can be defined as the set of stars meeting the condition:
\begin{equation}
\left\vert \mathbf{n}_{i}-\mathbf{n}_{0}\right\vert \leq\rho,\qquad i=1,...M.
\label{sa2}%
\end{equation}
One can define for each event its local orthonormal frame defined on the
basis:
\begin{align}
\mathbf{k}_{r}  &  =\mathbf{n}_{0},\label{sa3}\\
\mathbf{k}_{l}  &  =(-\sin l_{0},\cos l_{0},0),\nonumber\\
\mathbf{k}_{b}  &  =(-\sin b_{0}\cos l_{0},-\sin b_{0}\sin l_{0},\cos
b_{0}),\nonumber
\end{align}
where $\mathbf{k}_{l}(\mathbf{k}_{b})$ represent local directions of the
longitude (latitude). The local coordinates are defined as%
\begin{equation}
x_{i}=\mathbf{n}_{i}^{\prime}.\mathbf{k}_{l},\qquad y_{i}=\mathbf{n}%
_{i}^{\prime}.\mathbf{k}_{b};\qquad\mathbf{n}_{i}^{\prime}=\mathbf{n}%
_{i}-\mathbf{n}_{0}. \label{sa4}%
\end{equation}
We work with the following representations of the star positions
(Fig.\ref{faa3}a) inside the event circle:

\textit{i)} Two-dimensional (2D) positions $\{x_{i},y_{i}\}$ (Sect. \ref{dod}):
\begin{equation}
\{x_{i},y_{i}\};\quad x_{i}^{2}+y_{i}^{2}\leq\rho^{2},\quad i=1,...M
\label{sa5}%
.\end{equation}

\textit{ii)} Azimuthal positions\ $\left\{  \varphi_{i}\right\}  $
(Sect. \ref{aac}):
\begin{equation}
\left\{  \varphi_{i}\right\}  ;\quad-\pi<\varphi_{i}<\pi,\quad i=1,...M
\label{sa6}%
,\end{equation}
defined by $\{x_{i},y_{i}\}:$
\begin{equation}
x_{i}=r_{i}\cos\varphi_{i},\quad y_{i}=r_{i}\sin\varphi_{i},\quad r_{i}%
=\sqrt{x_{i}^{2}+y_{i}^{2}}. \label{sa7}%
\end{equation}
Here we use two sources of input data: (A) The simulated events generated by the Monte-Carlo (MC) code; Fig.\ref{faa1}%
a is an example of uniform generation of the star positions, and
(B) the real star events obtained from the Gaia catalog.
The methods for their analysis are described below. The final results of the
analysis follow from the comparison of the parameters and distributions
obtained from both sources of input data.

\subsection{Fourier analysis}

\label{aac} We start from the general formalism introduced in
\cite{Voloshin:1994mz,Poskanzer:1998yz}. The angular distribution
$P(\varphi)>0$ in $(-\pi,\pi)$ can be expressed as the Fourier series
\begin{align}
P(\varphi)  &  =\frac{1}{2\pi}\left(  1+2\sum_{n=1}^{\infty}v_{n}\cos\left[
n\left(  \varphi-\Psi_{n}\right)  \right]  \right)  ,\label{az1}\\
\int_{-\pi}^{\pi}P(\varphi)d\varphi &  =1,\nonumber
\end{align}
where the set of parameters $\left\{  v_{n},\Psi_{n}\right\}  $ define the
distribution. If we define the mean value of the function $f$ as%
\begin{equation}
\left\langle f\left(  \varphi\right)  \right\rangle \equiv\int_{-\pi}^{\pi
}P(\varphi)f\left(  \varphi\right)  d\varphi, \label{az4}%
\end{equation}
then the orthogonality of the terms in Eq. (\ref{az1}) implies:%
\begin{align}
v_{n}  &  =\left\langle \cos\left[  n\left(  \varphi-\Psi_{n}\right)  \right]
\right\rangle ,\label{az2a}\\
\tan\left(  n\Psi_{n}\right)   &  =\frac{\left\langle \sin\left(
n\varphi\right)  \right\rangle }{\left\langle \cos\left(  n\varphi\right)
\right\rangle }, \label{az2b}%
\end{align}
for any $n=1,2,3, ...$ . If we take event (\ref{sa6}) in which the
probability of $\varphi_{i}$ is proportional to $P(\varphi_{i})$\ and replace
the average value defined by Eq. (\ref{az4}) with the summation
\begin{equation}
\left\langle f\left(  \varphi\right)  \right\rangle _{M}\equiv\frac{1}{M}%
\sum_{k=1}^{M}f(\varphi_{k}), \label{az5}%
\end{equation}
then instead of Eqs. (\ref{az2a}) and (\ref{az2b}) we get%
\begin{align}
v_{n}(M)  &  =\left\langle \cos\left[  n\left(  \varphi-\Psi_{n}\right)
\right]  \right\rangle _{M},\label{az3a}\\
\tan\left(  n\Psi_{n}(M)\right)   &  =\frac{\left\langle \sin\left(
n\varphi\right)  \right\rangle _{M}}{\left\langle \cos\left(  n\varphi\right)
\right\rangle _{M}}. \label{az3b}%
\end{align}
Apparently, for any $n=1,2,3,...$ and $M\rightarrow\infty$ \ one can expect:%
\begin{equation}
\left\langle f\left(  \varphi\right)  \right\rangle _{M}\rightarrow
\left\langle f\left(  \varphi\right)  \right\rangle ,\qquad v_{n}%
(M)\rightarrow v_{n},\qquad\Psi_{n}(M)\rightarrow\Psi_{n}. \label{az6}%
\end{equation}
The Eqs. (\ref{az3a}) and (\ref{az3b}) do not have an unambiguous
solution, as illustrated in Fig. \ref{faa2}.
\begin{figure*}[t]
\centering\includegraphics[width=18cm]{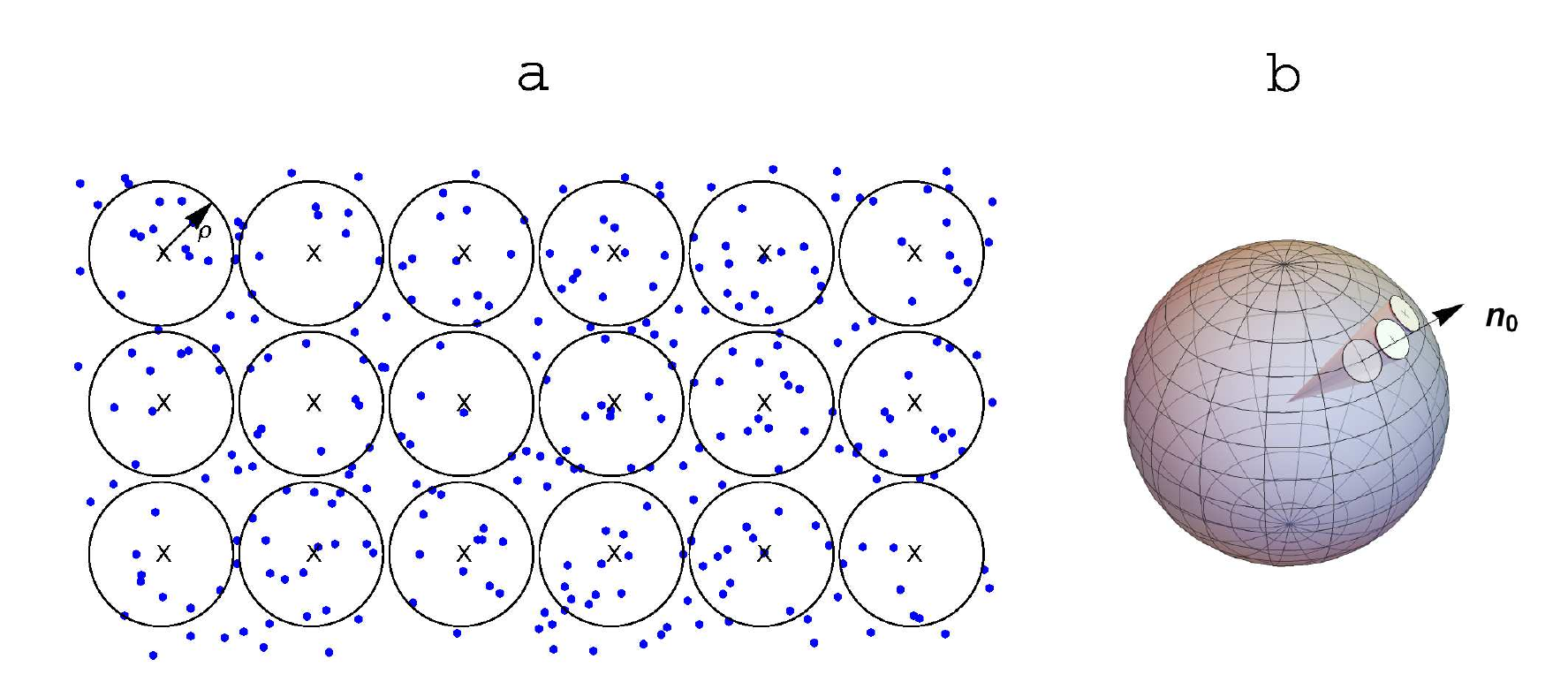}\caption{Patterns of the stars
inside the grid of circles (a). Position \textbf{n}$_{0}$ of the event
defines its local reference frame (b).}%
\label{faa1}%
\end{figure*}
\begin{figure*}[t]
\centering\includegraphics[width=14cm]{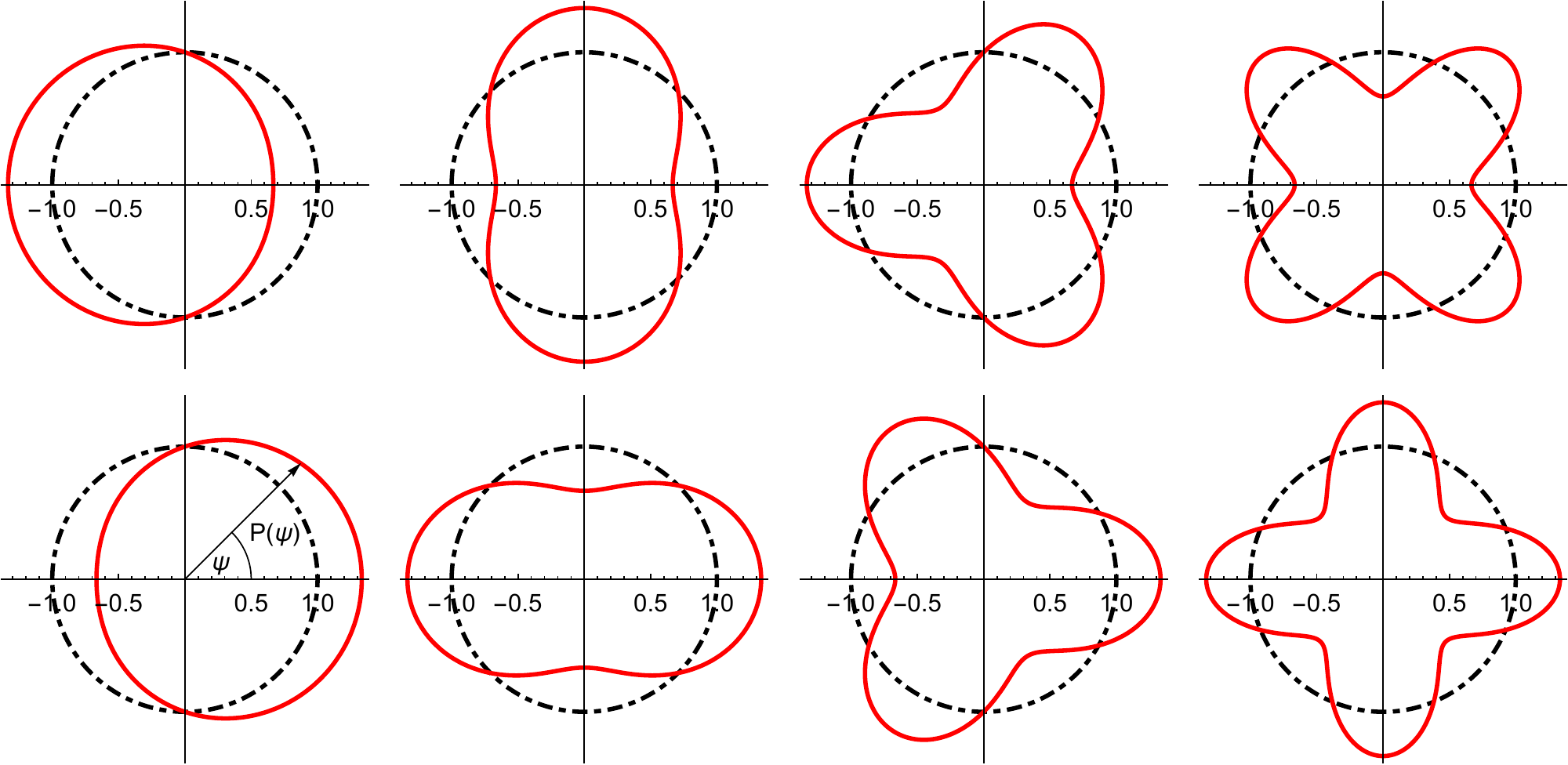}\caption{Examples of
deviations from uniform angular distribution defined by Eq.(\ref{az61}). The
red curves are their representation in polar coordinates $(P(\psi),\psi)$ as
indicated in the lower left panel. The panels from left to right correspond to
$n=1,2,3,4.$ The upper (lower) panels correspond to the negative (positive)
sign of \ $v_{n}=\mp1/6$. The black dot-dashed circles correspond to uniform
distribution $P(\psi)=1.$}%
\label{faa2}%
\end{figure*}

The panels represent distributions%
\begin{equation}
P(\psi)=1+2v_{n}\cos(n\psi),\qquad\psi=\varphi-\Psi_{n}. \label{az61}%
\end{equation}
The relation in Eq. (\ref{az3b}) gives the solutions%
\begin{equation}
n\Psi_{n}(M)=\tan^{-1}\left[  \frac{\left\langle \sin\left(  n\varphi\right)
\right\rangle _{M}}{\left\langle \cos\left(  n\varphi\right)  \right\rangle
_{M}}\right]  +k\pi. \label{az62}%
\end{equation}
The sign of the term $v_{n}\cos\left[  n\left(  \varphi-\Psi_{n}\right)
\right]  ~$\ in Eq. (\ref{az1}) can be controlled either by the sign of $v_{n}$ or
by the phase $n\Psi_{n}$. Apparently the change $v_{n}\rightarrow-v_{n}$ is
equivalent to $n\Psi_{n}\rightarrow n\Psi_{n}\pm k\pi,\quad k=1,2,3,...$  . 
Nevertheless in the present paper we analyze only $v_{n}^{2}(M)$ and not
$v_{n}(M)$ with the phase $\Psi_{n}(M)$. Therefore the sign ambiguity does not
play a role.

The relation in Eq. (\ref{az3a}) implies%
\begin{equation}
v_{n}(M)=\cos\left(  n\Psi_{n}\right)  \left\langle \cos\left(  n\varphi
\right)  \right\rangle _{M}+\sin\left(  n\Psi_{n}\right)  \left\langle
\sin\left(  n\varphi\right)  \right\rangle _{M}. \label{az7}%
\end{equation}
The relations%
\begin{equation}
\sin x=\frac{\tan x}{\sqrt{1+\tan^{2}x}},\qquad\cos x=\frac{1}{\sqrt
{1+\tan^{2}x}} \label{az8}%
\end{equation}
together with Eq. (\ref{az3b}) give%
\begin{align}
\sin\left(  n\Psi_{n}\right)   &  =\frac{\left\langle \sin\left(
n\varphi\right)  \right\rangle _{M}}{\sqrt{\left\langle \cos\left(
n\varphi\right)  \right\rangle _{M}^{2}+\left\langle \sin\left(
n\varphi\right)  \right\rangle _{M}^{2}}},\label{az9a}\\
\cos\left(  n\Psi_{n}\right)   &  =\frac{\left\langle \cos\left(
n\varphi\right)  \right\rangle _{M}}{\sqrt{\left\langle \cos\left(
n\varphi\right)  \right\rangle _{M}^{2}+\left\langle \sin\left(
n\varphi\right)  \right\rangle _{M}^{2}}}. \label{az9b}%
\end{align}
Inserting of these latter two expressions into Eq. (\ref{az7}) after some algebra
gives%
\begin{equation}
v_{n}^{2}(M)=\left\langle \cos\left(  n\varphi\right)  \right\rangle _{M}%
^{2}+\left\langle \sin\left(  n\varphi\right)  \right\rangle _{M}^{2}.
\label{az101}%
\end{equation}
%\textgreater
We note that $v_{n}^{2}$ does not depend on the choice of $k$ in \EMPH{Eq.} (\ref{az62}).
The last relation can be modified as%
\begin{equation}
v_{n}^{2}(M)=\frac{1}{M}\left[  1+\frac{2}{M}\sum_{1\leq k<l\leq M}%
\cos(n\varphi_{k}-n\varphi_{l})\right]  . \label{az11}%
\end{equation}
There are two extreme cases:

\EMPH{I)} All angles are in a narrow cone so that $\cos(n\varphi_{i}-n\varphi
_{j})\approx1 $; subsequently%
\begin{equation}
\sum_{1\leq k<l\leq M}\cos(n\varphi_{k}-n\varphi_{l})\approx\frac{M(M-1)}{2}
\label{az16}%
,\end{equation}
and%
\begin{equation}
v_{n}^{2}(M)\approx1. \label{az17}%
\end{equation}

\begin{figure*}[t]
\centering\includegraphics[width=14cm]{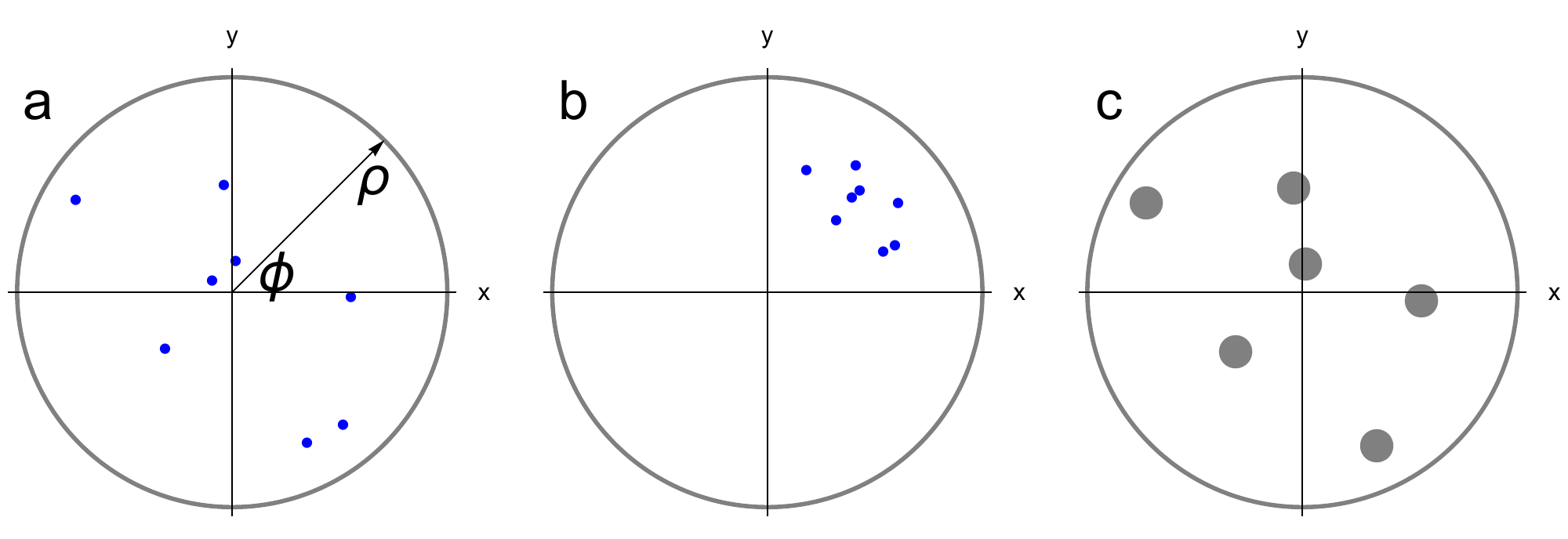}\caption{Examples of the
events generated by different algorithms: uniform distribution (a), clustering
$\lambda=1/3$ (b), and anti-clustering $\lambda=1/20$ (c).}%
\label{faa3}%
\end{figure*}

\EMPH{II)} All angles are regularly distributed on the circle, so $\varphi_{k}=2\pi
k/M$ (anti-clustering). If we define $q=\exp(i2\pi n/M)$, $n/M\neq
1,2,3,...,$\ then we get%
\begin{align}
\sum_{1\leq k<l\leq M}\cos(n\varphi_{k}-n\varphi_{l})  &  =\operatorname{Re}%
\sum_{1\leq k<l\leq M}q^{k-l}\label{az18}\\
&  =-\operatorname{Re}\frac{q\left[  1+M\left(  q-1\right)  -q^{M}\right]
}{\left(  q-1\right)  ^{2}}\nonumber\\
&  =M\operatorname{Re}\frac{q}{1-q}=-\frac{M}{2},\nonumber
\end{align}
which after inserting to Eq. (\ref{az11}) gives%
\begin{equation}
v_{n}^{2}(M)=0. \label{az19}%
\end{equation}
This result confirms an expectation that the regular distribution should not
generate asymmetry terms in Eq. (\ref{az1}).\newline In general, we can have
$N_{M}$ events of the same multiplicity $M$%
\begin{align}
-\pi\leq\theta_{j}-\Delta\leq\varphi_{1}^{j},..,\varphi_{M}^{j}\leq\theta
_{j}+\Delta\leq\pi, \quad j=1,..N_{M},\label{az20}
\end{align}
where $\theta_{j}$ can be different for various events. Subsequently, the average value
of $v_{n}^{2}(M)$ defined by Eq. (\ref{az11}) can be estimated as%
\begin{equation}
\left\langle v_{n}^{2}(M)\right\rangle =\frac{1}{M}\left[  1+\frac{2}{M}%
\sum_{1\leq k<l\leq M}\left\langle \cos(n\varphi_{k}^{j}-n\varphi_{l}%
^{j})\right\rangle \right]  , \label{az12}%
\end{equation}
where%
\begin{equation}
\left\langle \cos(n\varphi_{k}^{j}-n\varphi_{l}^{j})\right\rangle =\frac
{1}{N_{M}}\sum_{j=1}^{N_{M}}\cos(n\varphi_{k}^{j}-n\varphi_{l}^{j}).
\label{az12a}%
\end{equation}
For uniform distribution of $\varphi_{k}^{j}$ inside the intervals
of Eq. (\ref{az20}), and with sufficiently great $N_{M}$ , one can estimate this value by the
integral
\begin{align}
&  \left\langle \cos(n\varphi-n\psi)\right\rangle \label{az21}\\
&  =\frac{1}{4\Delta^{2}}\int_{-\Delta}^{\Delta}\int_{-\Delta}^{\Delta}\left(
\cos n\varphi\cos n\psi+\sin n\varphi\sin n\psi\right)  d\varphi
d\psi\nonumber\\
&  =\left(  \frac{\sin n\Delta}{n\Delta}\right)  ^{2}.\nonumber
\end{align}
Subsequently, for any $n=1,2,3,...$ we have%
\begin{equation}
\left\langle v_{n}^{2}(M)\right\rangle =\frac{1}{M}\left[  1+\left(
M-1\right)  \left(  \frac{\sin n\Delta}{n\Delta}\right)  ^{2}\right]
\geq\frac{1}{M}, \label{az22}%
\end{equation}
and therefore for $\Delta\rightarrow0$ (\textit{clustering}), we obtain%
\begin{equation}
\left\langle v_{n}^{2}(M)\right\rangle \rightarrow1, \label{az13}%
\end{equation}
which corresponds to \EMPH{case (I) of Eq. (\ref{az16}).} For $\Delta=\pi$ (\textit{uniform
distribution}) one obtains%
\begin{equation}
\left\langle v_{n}^{2}(M)\right\rangle =\frac{1}{M}. \label{az14}%
\end{equation}
To explain the practical meaning of the relation in Eq. (\ref{az13}), let us assume the
stars inside the circle in Fig.\ref{faa3}a are not distributed uniformly over
this circle, but are concentrated in some smaller circle, which is located
anywhere inside the greater one in Fig.\ref{faa3}b; their positions
$\left\{  \varphi_{1}...\varphi_{M}\right\}  $ inside the greater circle fill
up a narrower angular segment (statistically). Such a scenario is reflected in
Eqs.(\ref{az17}) and (\ref{az13}). In general there can be a mixture
\textit{uniform + clustering}, something between (\ref{az13}) and
(\ref{az14}):%
\begin{equation}
\frac{1}{M}<\left\langle v_{n}^{2}\left(  M\right)  \right\rangle <1.
\label{az141}%
\end{equation}

 Furthermore, the relation
\EMPH{in Eq. (\ref{az14})} corresponds to the uniform distribution, and for $M\rightarrow\infty$ gives%
\begin{equation}
\left\langle v_{n}^{2}(M)\right\rangle \rightarrow0, \label{az142}%
\end{equation}
which is the correct result for uniform distribution defined by Eq. (\ref{az1}).
But why does a finite $M$ generate $\left\langle v_{n}^{2}(M)\right\rangle >0$
even for uniform generation? The reason is that the event of finite
multiplicity, for example $M=3$ of random stars, is usually better described
with the use of higher harmonics. For increasing $M$ \ the population becomes denser and more symmetric in terms of $\left\{  \varphi_{1}%
...\varphi_{M}\right\}  $, in accordance with Eq. (\ref{az142}).

Here we also use the function $\Theta_{n}$ defined as%

\begin{equation}
\Theta_{n}(M)=M\left\langle v_{n}^{2}(M)\right\rangle . \label{az143}%
\end{equation}

For illustration we present the toy examples of simulation:

\textit{a) Uniform distribution}

\begin{figure*}[t]
\centering\includegraphics[width=17cm]{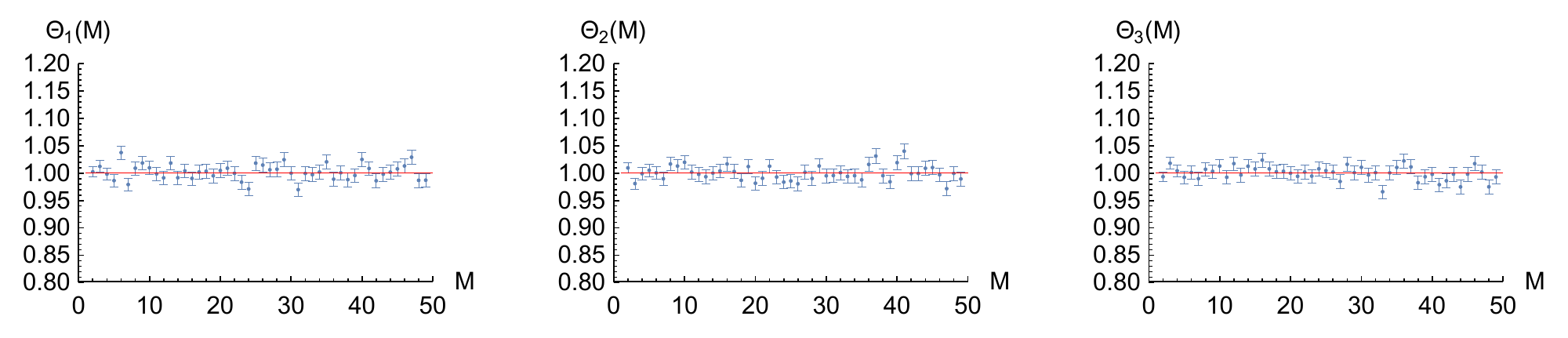}
\includegraphics[width=17cm]{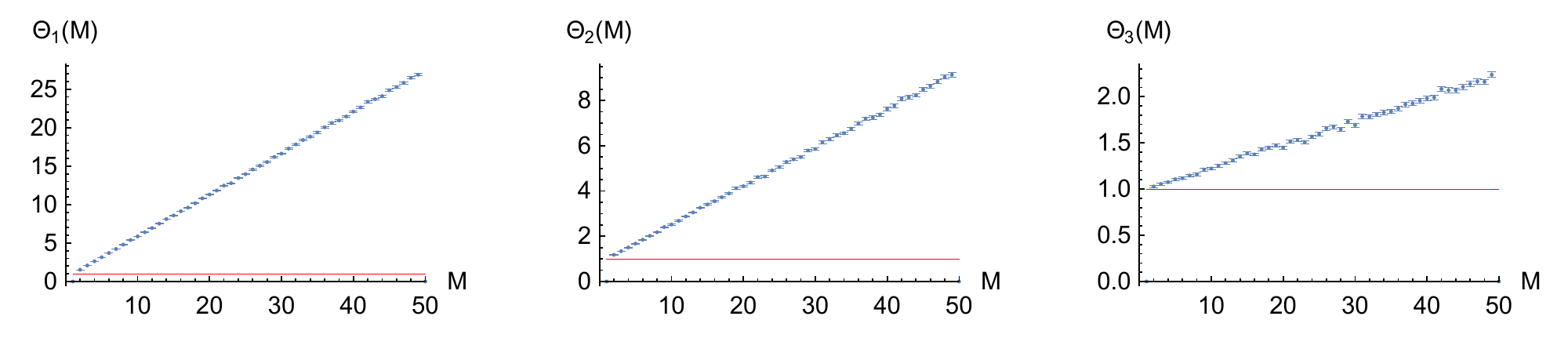}
\includegraphics[width=17cm]{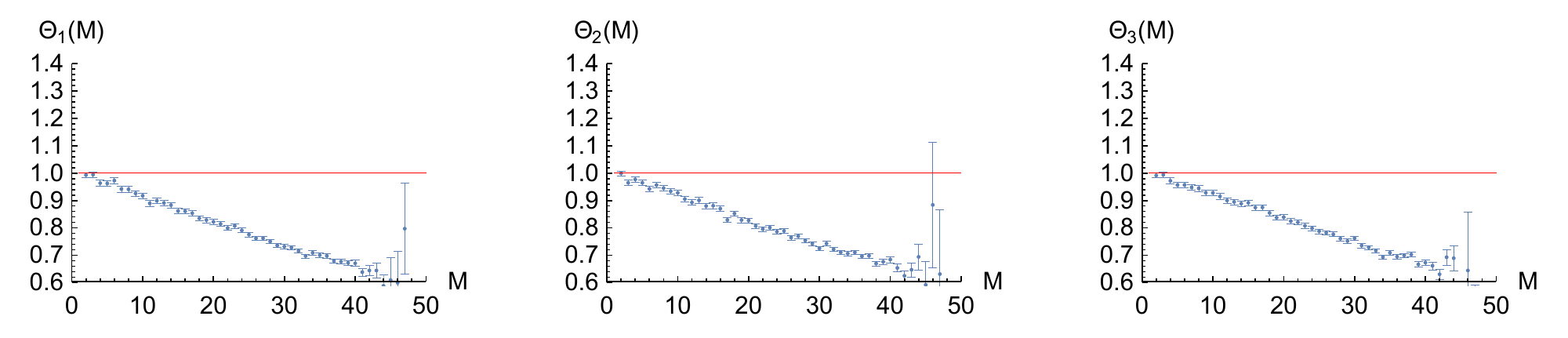}\caption{The functions $\Theta_{n}(M)$,
$n=1,2,3$\ for Monte-Carlo events in the scenario of uniform distribution
(upper panels). The remaining panels show clustering $\lambda=1/3$ (middle
panels) and anti-clustering scenarios $\lambda=1/20$ (lower panels). Each
multiplicity bin is generated by $N_{M}=6000$ events. The red lines correspond
to the expected dependence for uniform distribution, $\Theta_{n}(M)=1$ (Eq.
(\ref{az14})).}%
\label{faa4}%
\end{figure*}

We generate uniform sets of stars inside the circle of radius $\rho,$ like the
event in Fig.\ref{faa3}a.
For each event we use the relation in Eq. (\ref{az101}) to calculate
$v_{n}^{2}(M)$ for $n=1,2,3$. The functions
\begin{equation}
\Theta_{n}(M)=M\left\langle v_{n}^{2}(M)\right\rangle =\frac{M}{N_{M}}%
\sum_{k=1}^{N_{M}}v_{n,k}^{2}(M), \label{az15}%
\end{equation}
where $N_{M}$\ is the number of events of multiplicity $M,$ are displayed in
the upper panels of \ Fig. \ref{faa4}. The resulting lines apparently satisfy
Eq. (\ref{az14}). We note that throughout the paper, the error bars, if plotted, indicate only
statistical errors. 

\textit{b) Clustering}

In a first step, we uniformly generate stars inside a smaller circle of radius
$\delta.$ This circle is randomly located inside the greater circle of
radius $\rho$, and we define $\lambda=\delta/\rho$ (Fig.\ref{faa3}b). The
functions $\Theta_{n}(\rho,\delta,M)$ are calculated with the use of
Eq. (\ref{az15}), equally as in the previous case. The results are displayed in
the middle panels of \ Fig. \ref{faa4}.

\textit{c) Anti-clustering}

We generate spots of the radius $\delta$ inside the circle of radius $\rho,$
giving the ratio $\lambda=\delta/\rho$ (Fig.\ref{faa3}c). The MC algorithm is
the same as for uniform distribution, but with the additional constraint that the
spots must not overlap. If two spots in uniform generation overlap, one of
them is excluded. In other words, there is a rule that the distance between any two
stars is greater than $2\delta$. The corresponding functions (\ref{az15}) are
displayed in the lower panels of Fig. \ref{faa4}. Why do the curves decrease?
Obviously, a denser population of spots inside the circle generates a more
regular arrangement of $\varphi_{k}$, closer to \EMPH{case (II)} above, so the
function $\Theta_{n}(M)$ will tend to approach the minimum given by
Eq. (\ref{az19}). The curves in the figure are \EMPH{linear for}
a small $\lambda$ and $M\ll M_{\max}\lesssim1/\lambda^{2}$
\EMPH{and one can approximate as}
\begin{equation}
\Theta_{n}(\rho,\delta,M)\approx1-3.36\lambda^{2}\left(  M-1\right)
\label{az25}%
,\end{equation}
for $n=1,2,3.$

Generally, for both algorithms mentioned above, the change of scale
$\rho\rightarrow k\rho$, $\delta\rightarrow k\delta$ does not change
distribution of angles in the event $\left\{  \varphi_{1}...\varphi
_{M}\right\}  $ and $\left\langle v_{n}^{2}(M)\right\rangle $ defined by
(\ref{az12}); therefore%
\begin{equation}
\Theta_{n}(\rho,\delta,M)=\Theta_{n}(\lambda,M);\qquad\lambda=\delta/\rho.
\label{az25a}%
\end{equation}
These algorithms are simple examples; one could think up other ones.

We summarize the main results of this section as follows.

\textit{i)} The uniform field (like the events in Fig.\ref{faa1}a) generate
the dependence $\Theta_{n}(M)=1$ for any $n=1,2,3,... $ . The non-uniform
distributions violate this rule. We have shown the examples, which generate
relations $\Theta_{n}(M)>1$ (clustering) and $\Theta_{n}(M)<1$ \ (anti-clustering).

\textit{ii) }In this way the functions $\Theta_{n}(M)$ give important
information about the statistical character of the patterns, but the numbers
$v_{n}^{2}(M)$ calculated only from a single event do not offer a significant amount of information.

The functions $\Theta_{n}(M)$ are used in Sect.\ref{ata} for analysis and
classification of the sets of real star events.

\subsection{Distributions of angular distances}

\label{dod} Inside event (\ref{sa5}) we define the angular distances:%
\begin{equation}
x_{ij}=\left\vert x_{i}-x_{j}\right\vert ,\quad y_{ij}=\left\vert y_{i}%
-y_{j}\right\vert ,\quad d_{ij}=\sqrt{x_{ij}^{2}+y_{ij}^{2}}. \label{az31a}%
\end{equation}
We also define the parameters characterizing the dimension of a triplet of stars:%
\begin{align}
x_{ijk}  &  =\frac{x_{ij}+x_{jk}+x_{ki}}{2},\qquad y_{ijk}=\frac{y_{ij}%
+y_{jk}+y_{ki}}{2},\label{az31b}\\
d_{ijk}  &  =\frac{2}{3}\sqrt{d_{ij}^{2}+d_{jk}^{2}+d_{ki}^{2}},\qquad
i,j,k=1,...M.\nonumber
\end{align}
Let us consider a \textit{uniform} field of stars (Fig.\ref{faa1}a). The example
of a single event (\ref{sa5}) is displayed in Fig. \ref{faa3}a. For the set of
events we can calculate probability distributions of the parameters defined
above. The distributions satisfy:

1) The shape of the (normalized) distribution of $y_{ij}$\ is the same as that of
$x_{ij}$; similarly for $x_{ijk}$\ and $y_{ijk}$. Therefore for the uniform
events, there are four different distributions,
\begin{equation}
P_{1}(x_{ij},\rho),\quad P_{2}(d_{ij},\rho),\quad P_{3}(x_{ijk},\rho),\quad
P_{4}(d_{ijk},\rho) \label{az31c}%
.\end{equation}

2) The shapes of these distributions do not depend on multiplicity. \ Since
the numbers $x_{j}$ are independent, the distribution of $x_{jk}$ is the same
for any $j,k$. Increasing $M$ means only greater density, that is, greater numbers of
$x_{j}$ and $x_{jk}$, but proportion between different scales of $x_{jk}$ does
not change. The same argument holds for the distribution of all the parameters
(\ref{az31a}),(\ref{az31b}).

\begin{figure*}[t]
\centering\includegraphics[width=16cm]{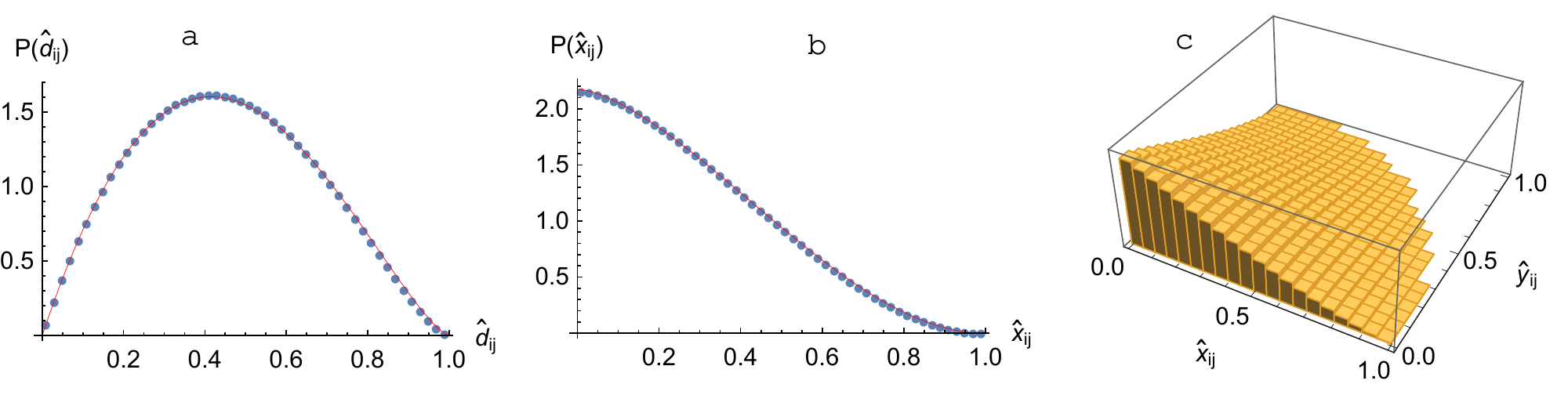}
\includegraphics[width=16cm]{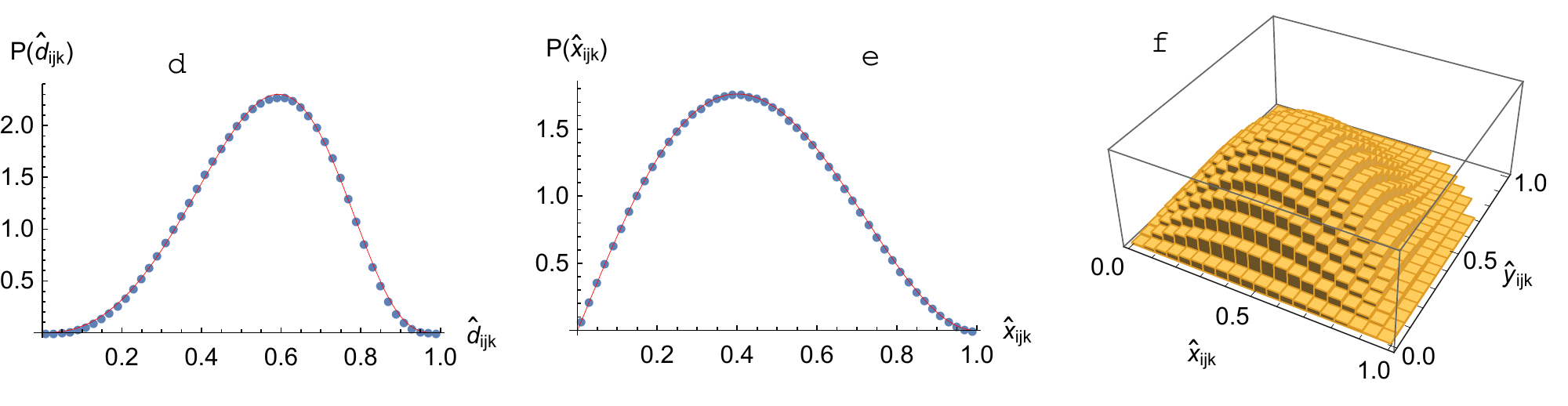}\caption{Distributions of angular
distances $P(\hat{x}_{\alpha})$ of uniformly generated stars (points) fitted
by the red curves (Eq. (\ref{az32})). Corresponding distribution $P(\hat{x}_{\alpha
},\hat{y}_{\alpha})$ is on the right. Variables in panels are defined by
relations (\ref{az31a}) - (\ref{az31}). The Monte-Carlo statistics is
represented by $2.5\times10^{6}$ events of multiplicity $M=5.$}%
\label{faa5}%
\end{figure*}

3) Obviously, there is the similarity $P(\Delta,\rho)\sim P(k\Delta,k\rho)$
for the distributions above, $\Delta=x_{\alpha},y_{\alpha},d_{\alpha}$. If we
rescale the distance parameters as%
\begin{equation}
\hat{x}_{\alpha}=\frac{x_{\alpha}}{2\rho},\quad\hat{y}_{\alpha}=\frac
{y_{\alpha}}{2\rho},\quad\hat{d}_{\alpha}=\frac{d_{\alpha}}{2\rho}\label{az31}%
,\end{equation}
then one can check that $0\leq\hat{\xi}\leq1$ for $\hat{\xi}=\hat{x}_{\alpha
},\hat{y}_{\alpha},\hat{d}_{\alpha}$ and $\alpha=ij$ or $\alpha=ijk.$
The corresponding normalized MC distributions $P\left(  \hat{\xi}\right)  $ are
together with the corresponding 3D plots $P\left(  \hat{x}_{\alpha},\hat
{y}_{\alpha}\right)  $ displayed in Fig. \ref{faa5}.
These distributions can be approximated by the function%
\begin{equation}
P\left(  \hat{\xi}\right)  =\frac{\Gamma\left(  1/a+b+1\right)  }%
{\Gamma\left(  1/a+1\right)  \Gamma\left(  b+1\right)  }\hat{\xi}^{a}\left(
1-\hat{\xi}^{c}\right)  ^{b},\quad\int_{0}^{1}P\left(  \hat{\xi}\right)
d\hat{\xi}=1,\label{az32}%
\end{equation}
where the parameters $a,b,c$ optimized by the fits in the whole region
$0<\hat{\xi}<1$ are listed in Table \ref{tbl1}. \begin{table}[ptb]
\caption{Parameters of distributions (\ref{az32}). }%
\label{tbl1}
\centering%
\begin{tabular}
[t]{|c|c|c|c|}\hline
variable & $a$ & $b$ & $c$\\\hline
$\hat{x}_{ij}$ & $0$ & $2$ & $1.551$\\\hline
$\hat{d}_{ij}$ & $1$ & $1.411$ & $1$\\\hline
$\hat{x}_{ijk}$ & $1.032$ & $2.113$ & $1.573$\\\hline
$\hat{d}_{ijk}$ & $2.349$ & $3.760$ & $3.709$\\\hline
\end{tabular}
\end{table}The universal plots in the figure, after rescaling $\hat{\xi
}\rightarrow\Delta=2\rho\hat{\xi}$ ,  will serve in the following section as the templates
for comparison with the real events of angular radius $\rho$. We must point out
that the functions (\ref{az32}) with the parameters in the table are only
approximations of the MC distributions.
The very good agreement in Fig. \ref{faa5} is due to
`flexibility' of this parameterization, which for $a,b,c\geq0$ satisfies
needed boundary conditions: $P\left(  \hat{\xi}\right)  \rightarrow0$ for
$\hat{\xi}\rightarrow0$ and $\hat{\xi}\rightarrow1$. The term with the
$\Gamma-$functions provides normalization.
Despite the simple MC algorithm for the definition of the
distributions $P\left(  \hat{\xi}\right)  $, we did not succeed in expressing
their exact form in terms of the known standard or special functions.
However, in principle we can calculate them with arbitrary precision,
which is needed for the data analysis. We refer to
the following rules of the MC technique. The statistical error of a simulated
distribution in the $k-$th bin is $\approx1/\sqrt{n_{k}}$, where $n_{k}$\ is
bin population. For an accurate analysis, this error should be much less than
the statistical error in the corresponding bin of the real data. In other words, the
number of simulated events should be substantially greater than the number of
the corresponding data events. In general, the precision of simulated
distributions increases with the number of generated events. 

\section{Application to the Gaia mission data}

\label{ata}
\begin{figure}[t]
\includegraphics[width=8.4cm]{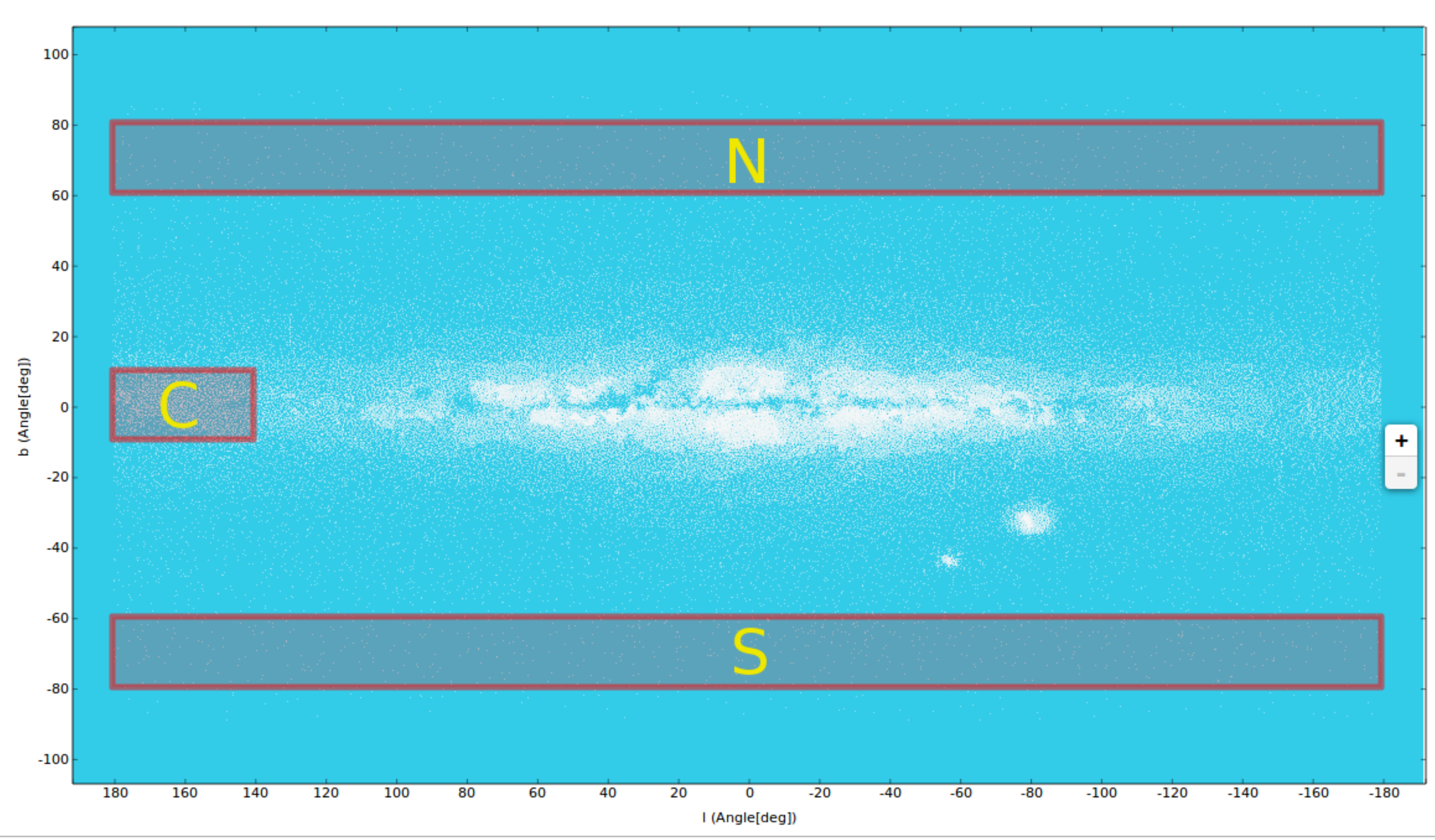}\newline%
\begin{tabular}
[c]{|c|c|c|c|c|}\hline
& area: $l\times b$ & $\rho\lbrack\deg]$ & $\left\langle M\right\rangle $ &
$N_{e}$\\\hline
C & $\left\langle 140,180\right\rangle \times\left\langle -10,10\right\rangle
$ & $0.005$ & $2.60$ & $6962671$\\\hline
N & $\left\langle -180,180\right\rangle \times\left\langle 60,80\right\rangle
$ & $0.02$ & $3.26$ & $1437067$\\\hline
S & $\left\langle -180,180\right\rangle \times\left\langle
-80,-60\right\rangle $ & $0.02$ & $3.30$ & $1448647$\\\hline
\end{tabular}
\caption{Analyzed regions in the Gaia catalog, where $\rho$ is angular radius
of the events, $M$ is their multiplicity and $N_{e}$ is the number of events.
Analysis is done only for events $2\leq M\leq15.$}%
\label{faa9}%
\end{figure}

The simulations described above have been applied to the analysis
of the data from the recent Gaia catalog DR1 \cite{gaia2}. We present the
results from the regions marked in Fig.\ref{faa9}.
In the present paper we are starting this analysis from the
simplest case, from the events of a small multiplicity. \EMPH{This condition refers}
to the small but different event radii in regions C and N\&S
(table in Fig.\ref{faa9}). In general, the scale of the possible structure
violating uniformity should be less than the event radius $\rho$.

\begin{figure*}[t]
\centering\includegraphics[width=18cm]{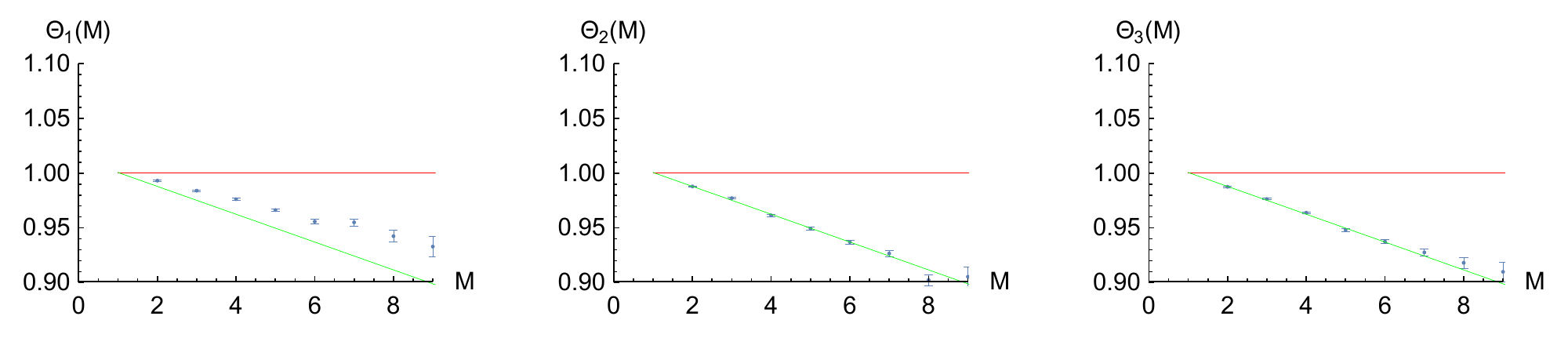}
\includegraphics[width=18cm]{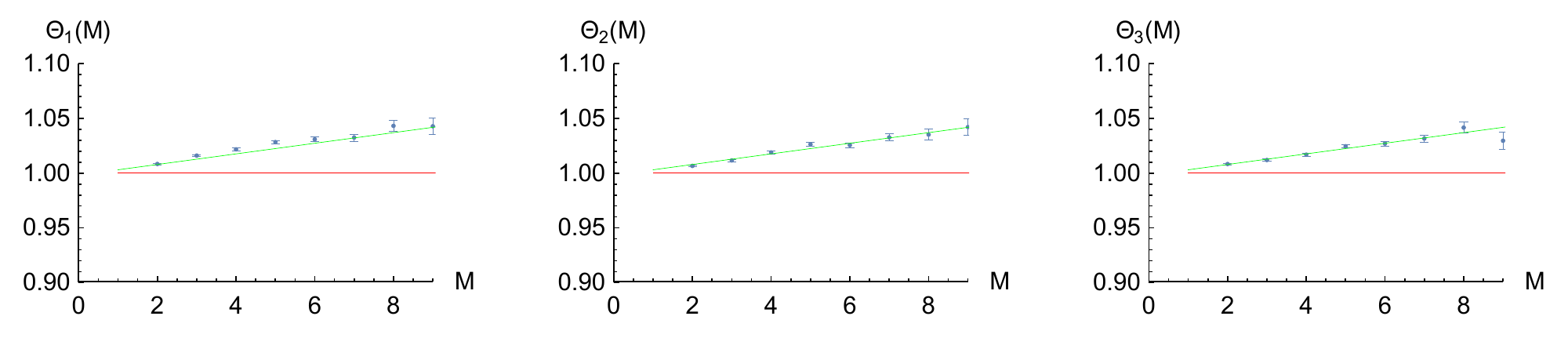}\caption{The functions $\Theta_{n}(M)$,
$n=1,2,3$\ for events in the area C (upper panels) and N\&S (lower panels).
The green line is a linear fit of data taken in panel $\Theta_{2}$ .}%
\label{faa6}%
\end{figure*}

First, we applied the Fourier analysis described in Sect. \ref{aac}. The
corresponding functions $\Theta_{n}(M)$ are shown in Fig. \ref{faa6}. The upper
part corresponds to a dense \EMPH{region} C and very similar results can be obtained
over other regions at the galactic plane.
The result qualitatively corresponds to the scenario of
anti-clustering simulated in the lower panels of Fig. \ref{faa4}, however the slope
of $\Theta_{1} $ appears to differ from the slopes of $\Theta_{2}$ and
$\Theta_{3}.$ The lower panels correspond to sparse \EMPH{region} N\&S and the slopes
$\Theta_{n}(M)$ suggest the presence of clustering.

\begin{figure*}[t]
\centering\includegraphics[width=16cm]{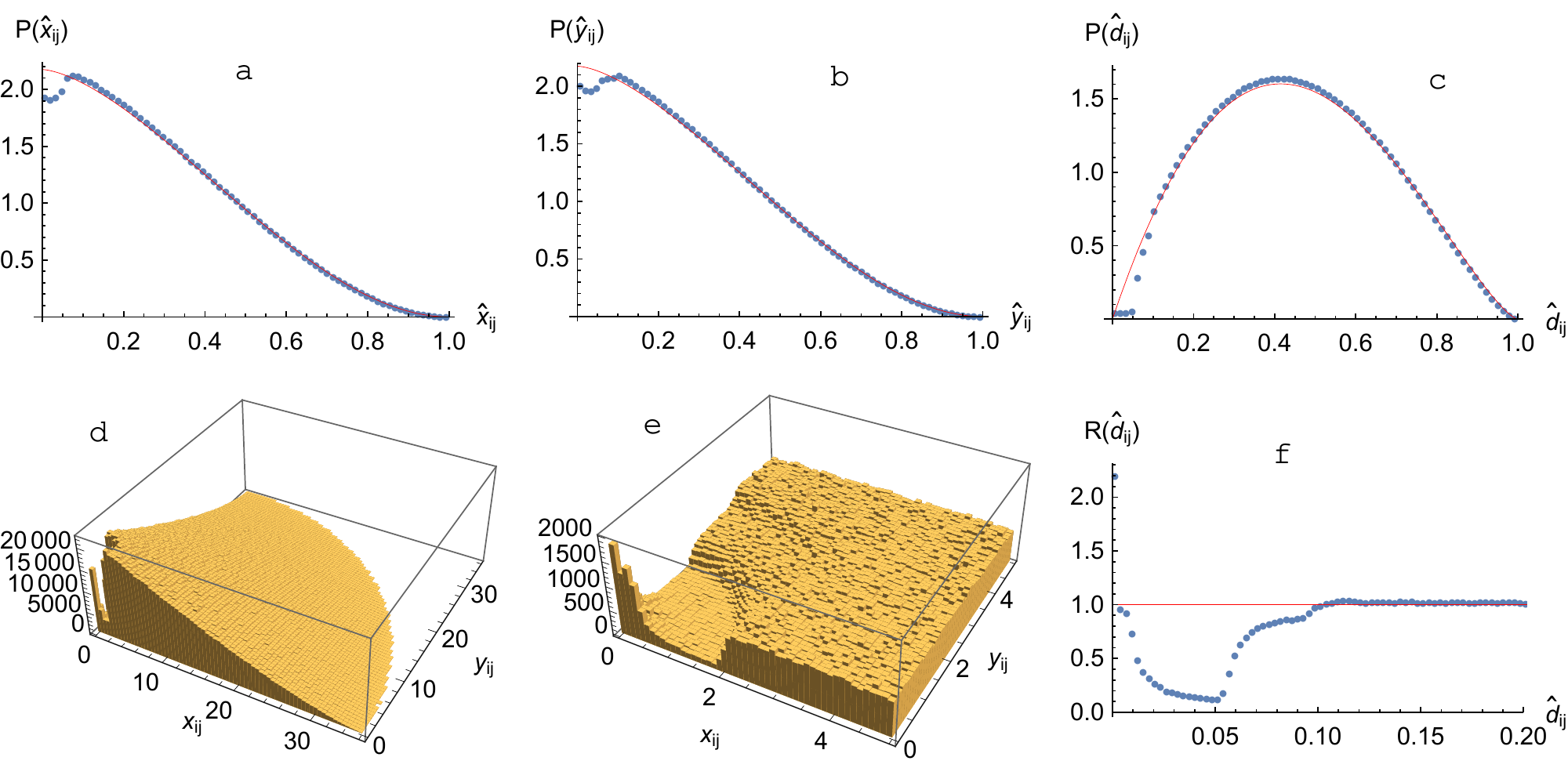}\caption{Distributions of
angular distances in region C for all $G$. The blue points in the panels
(a,b,c) represent the data on $\hat{x}_{ij},\hat{y}_{ij},\hat{d}_{ij}$ and the
red curves are the result of a Monte-Carlo simulation. Panel (f) is the ratio of
data to the simulation from panel (c). The panels (d,e) represent 3D plots of
distances $x_{ij},y_{ij}$; their unit$\ $is $1^{\prime\prime}$. Panel (e)
displays the region of small distances in higher resolution.}%
\label{faa7}%
\end{figure*}

To better understand these results, we have carried out a further analysis with the
method described in Sect. \ref{dod}. In Fig. \ref{faa7}
we show distributions of angular distances studied in \EMPH{region} C,
where we work with the events of radius $\rho=0.005\deg=18^{\prime\prime}$,
which means $d_{ij\max}=x_{ij\max}=y_{ij\max}=36^{\prime\prime}$, as seen in panel (d). For comparison with simulation of uniform events we used the
variables $\hat{x}_{ij},\hat{y}_{ij},\hat{d}_{ij}$ defined in Eq. (\ref{az32}).
The MC curves are the same as in panels (a,b) of Fig. \ref{faa5}. For
$\hat{d}_{ij}\gtrsim0.11$ (or equivalently $d_{ij}\gtrsim4^{\prime\prime}$)
the data agree perfectly with the MC simulation. However, the\ perfect
agreement expected for uniform events is violated for $d_{ij}\lesssim
4^{\prime\prime}$. \ In fact we observe a 2D representation of the effect
reported in \cite{gaia3} (Sect. 4.4.1., Fig.17), which is due to reduced
resolution of two sources in the same region of separation. Reduced efficiency
at small distances imitates the anti-clustering scenario. If we take
$2\delta=d_{ij\min}\approx2^{\prime\prime}$, then with the use of relation
(\ref{az25}) one can predict the slope of \ the corresponding function
$\Theta_{n}(M)$ in the lower panels of Fig. \ref{faa6} as%
\begin{equation}
k\approx3.36\lambda^{2}=3.36\left(  \frac{\delta}{\rho}\right)  ^{2}%
\approx0.01,\label{az33}%
\end{equation}
which gives a very reasonable agreement.

\begin{figure*}[t]
\centering\includegraphics[width=16cm]{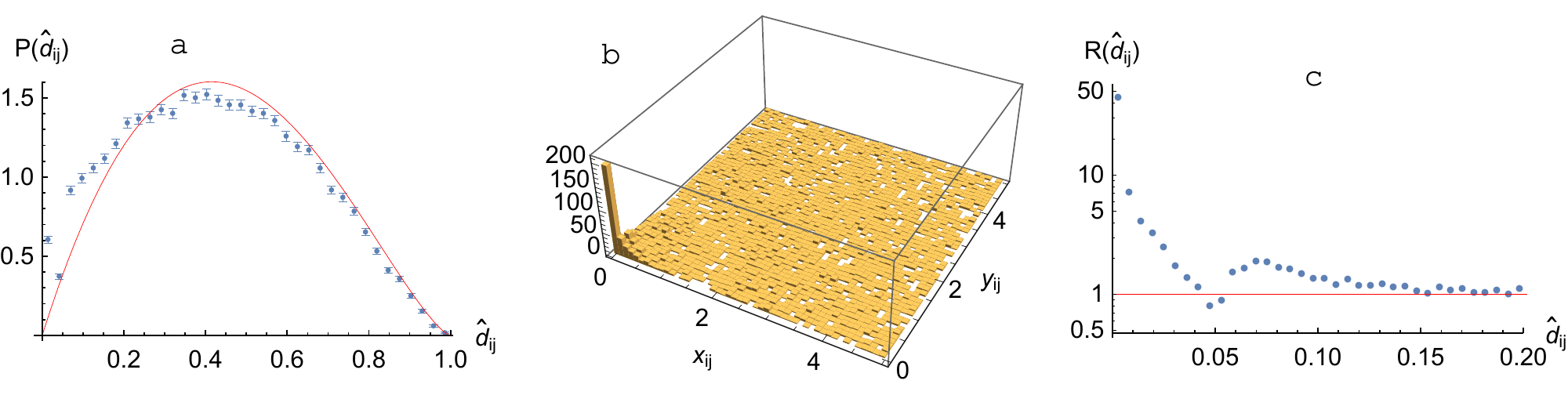}\caption{Distributions of
angular distances in region C for $G\leq15$. The blue points in panel
(a) represent the data on $\hat{d}_{ij}$ and the red curve is the result of
a Monte-Carlo simulation. Panel (b) represents a 3D plot of distances
$x_{ij},y_{ij}$, their unit$\ $is $1^{\prime\prime}$. Panel (c) is the
ratio of data to simulation from panel (a).}%
\label{faa8}%
\end{figure*}

In the figure we also observe a peak at small separation $d_{ij}%
\lesssim1^{\prime\prime}$. In \cite{gaia3} such a peak is observed only in
the sparse \EMPH{region} and is interpreted as the presence of binary stars. A
simulation described in the same paper suggests that reduced efficiency at
$d_{ij}\lesssim4^{\prime\prime}$ correlates with fainter magnitudes $G$ in the
\EMPH{region}. This is in agreement with our Fig. \ref{faa8}, where the sources with
$G>15$ mag are excluded and as a result the resolution dip is reduced.

\begin{figure*}[t]
\centering\includegraphics[width=16cm]{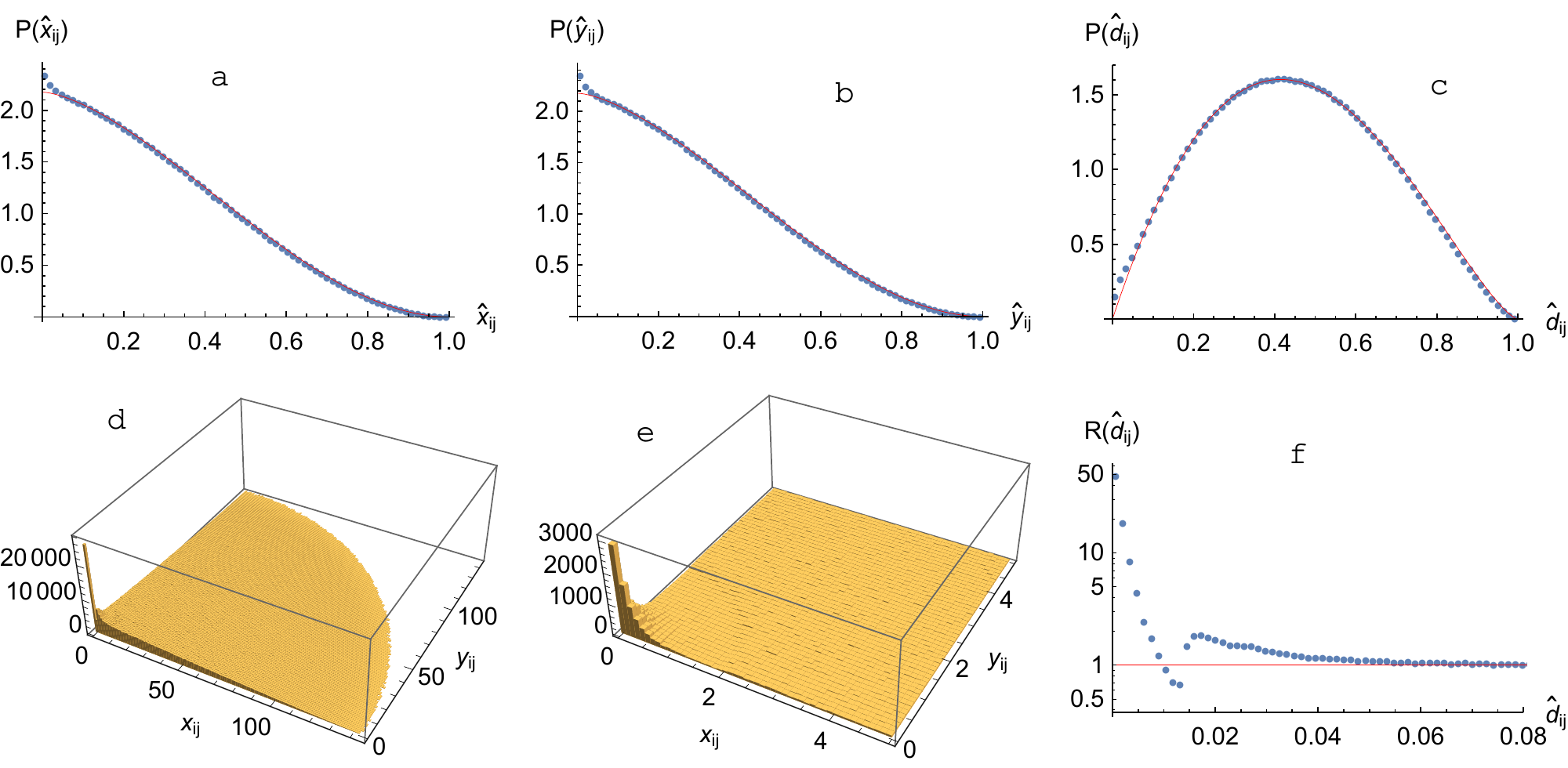}\caption{Distributions of
angular distances in the region N\&S for all $G$. The blue points in the
panels (a,b,c) represent the data on $\hat{x}_{ij},\hat{y}_{ij},\hat{d}_{ij}$
and the red curves are result of Monte-Carlo simulation. The panel (f) is the
ratio of data to simulation from panel (c). The panels (d,e) represent 3D plots
of distances $x_{ij},y_{ij}$; their unit $\ $is $1^{\prime\prime}$. Panel (e)
displays the region of small distances in higher resolution.}%
\label{faa10}%
\end{figure*}

\begin{figure*}[t]
\centering\includegraphics[width=16cm]{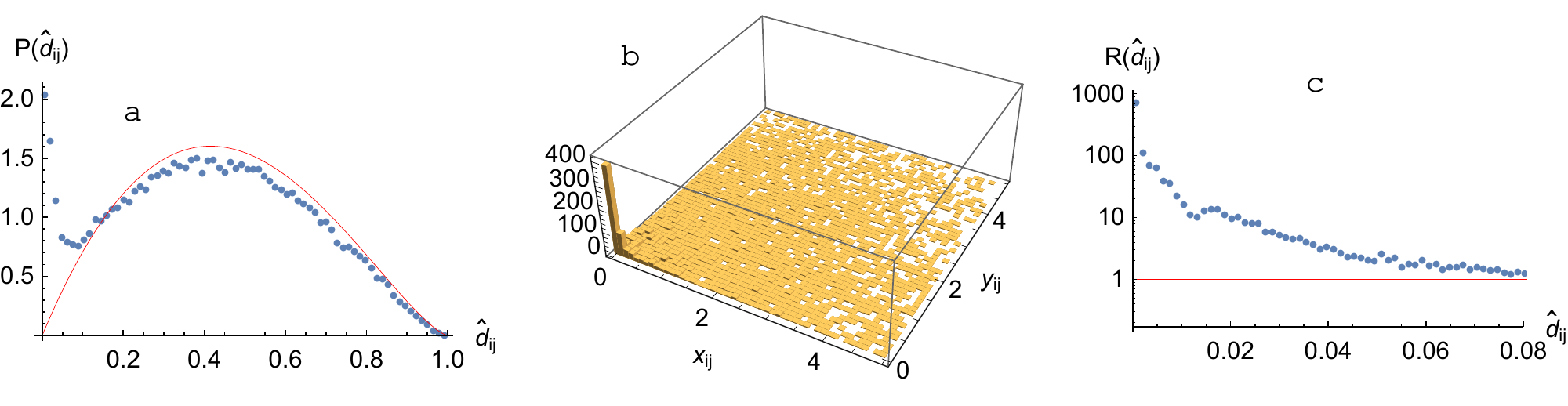}\caption{Distributions of
angular distances in the region N\&S for $G\leq15$. The blue points in
panel (a) represent the data on $\hat{d}_{ij}$ and the red curve is the result of
a Monte-Carlo simulation. Panel (b) represents a 3D plot of distances
$x_{ij},y_{ij}$; their unit$\ $is $1^{\prime\prime}$. Panel (c) is the
ratio of data to simulation from panel (a).}%
\label{faa11}%
\end{figure*}

A similar analysis for region N\&S is demonstrated in Figs.\ref{faa10} and
\ref{faa11}. Now we work with the events of radius $\rho=0.020\deg
=72^{\prime\prime}$, which means $d_{ij\max}=x_{ij\max}=y_{ij\max}%
=144^{\prime\prime}$. In the first figure, for $\hat{d}_{ij}%
(d_{ij})\gtrsim0.06(8.6^{\prime\prime})$ we see that the data agree perfectly with the MC
simulation.
Comparison of panel (f) from both Figs. \ref{faa7} and \ref{faa10} reveals that
in the sparse \EMPH{region} the efficiency drop is less pronounced. Further, in the
latter figure, apart from the pronounced peak at $d_{ij}\lesssim1^{\prime\prime}$,
we observe a clear excess of pairs separated by $\hat{d}_{ij}(d_{ij}%
)\lesssim0.06(8.6^{\prime\prime})$. In Fig.\ref{faa11}
we display results for brighter stars, $G\leq15$. The panels are
different representations of the very clear excess of pairs with a small
separation. In panel (a) we observe a pronounced peak at small distances on
the background, represented by the red curve in panel (a). We note that
the data and curve are equally normalized for $0<\hat{d}_{ij}<1$. The
background is generated by uniform distribution of the star pairs. This
background also naturally involves close pairs -- but only double stars without
the gravitational bond. The peak must be the result of some additional rule,
which makes\ the close pairs more frequent. We interpret this surplus as the
presence of the binary star systems (with gravitational bond). Panel (c)
displays the excess most explicitly, as statistical ratio binaries/background.
Further, we can observe some correspondence between Figs. \ref{faa8}c and
\ref{faa11}c. For instance the positions of local minima of $\hat{d}_{ij}$
$\approx0.05 (0.0125)$  correspond to $d_{ij}=2\rho\hat{d}_{ij}$, which gives
$d_{ij}\approx1.8^{\prime\prime}$ for both event radii $\rho=18^{\prime\prime
} (72^{\prime\prime})$. This correspondence confirms that the results of
the analysis should not be \mbox{sensitive to $\rho.$}

\begin{figure*}[t]
\centering\includegraphics[height=4cm]{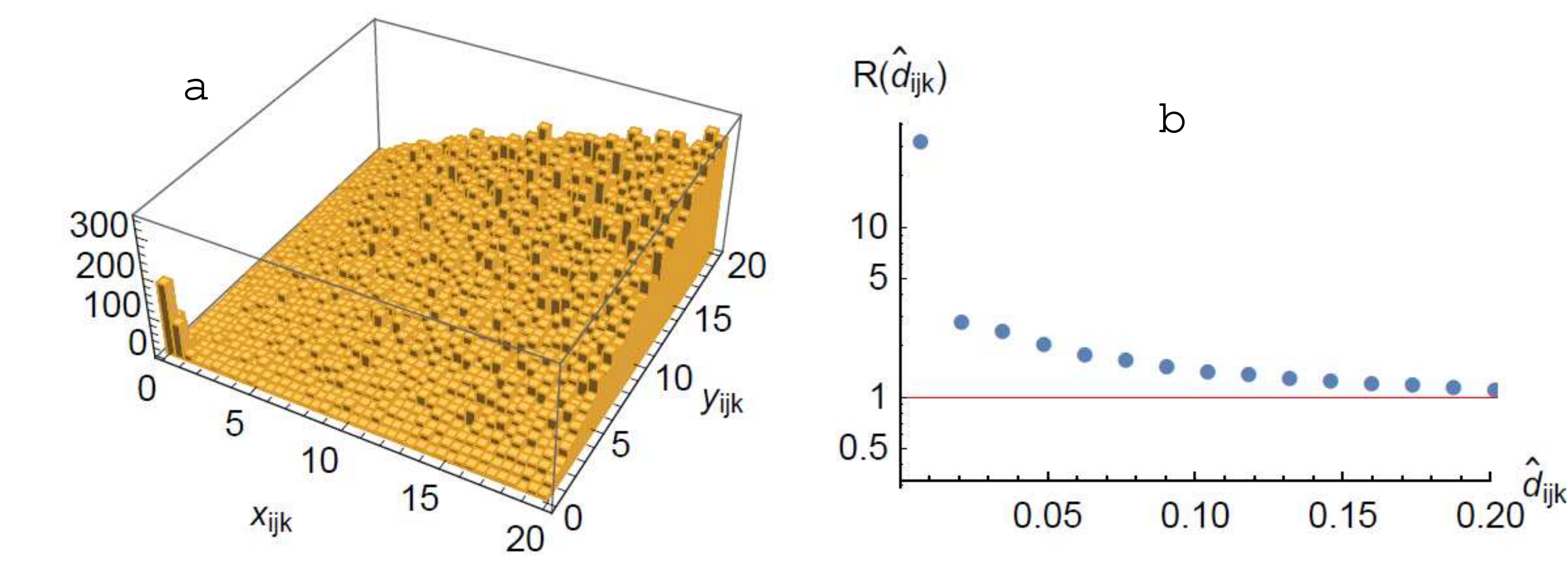}\includegraphics[height=4cm]{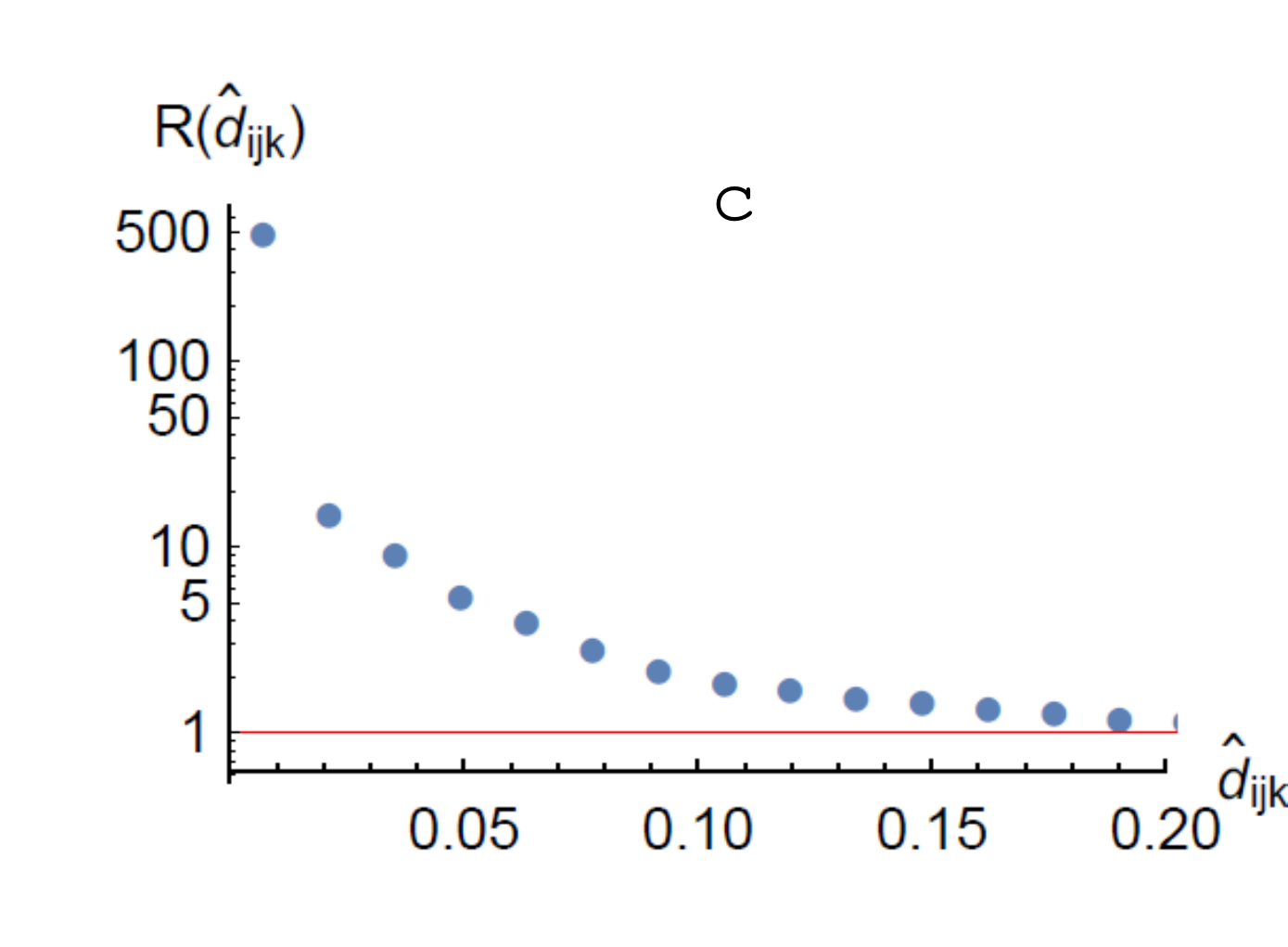}
\caption{3D
plot of angular ternary distances $x_{ijk},y_{ijk}$ (unit$\ $is $1^{\prime
\prime}$) in the region N\&S for all $G$ (a) and ratios of measured
distribution of relative distances $\hat{d}_{ijk}$ to the interpolation of
corresponding Monte-Carlo distribution for all $G$ (b) and $G\leq18$ mag (c).}%
\label{faa12}%
\end{figure*}

In a similar way we have analyzed distributions of the parameters
(\ref{az31b}) related to the triplets of stars in the \EMPH{region} N\&S. With the
condition $M\geq3$\ we have used the same events, $\rho=0.020\deg
=72^{\prime\prime}$, which means $d_{ijk\max}=x_{ijk\max}=y_{ijk\max
}=144^{\prime\prime}$. The main results are presented in Fig.\ref{faa12}.
Again, we observe a peak in the region of smaller triplet
distances; however the excess is broader and less dependent on $G$ than for
binaries. We interpret it as the presence of bounded ternary star systems.
Panels (b,c) display the statistical ratio ternaries/background.

The occurrence of the bounded star systems is a manifestation of the
clustering scenario. This scenario was suggested already by the function
$\Theta_{n}(M)$ in the lower panels of Fig.\ref{faa6} and followed from the
Fourier analysis.

\subsection{Gravitational microlensing}

\label{wdn}
\begin{figure*}[t]
\centering\includegraphics[width=14cm]{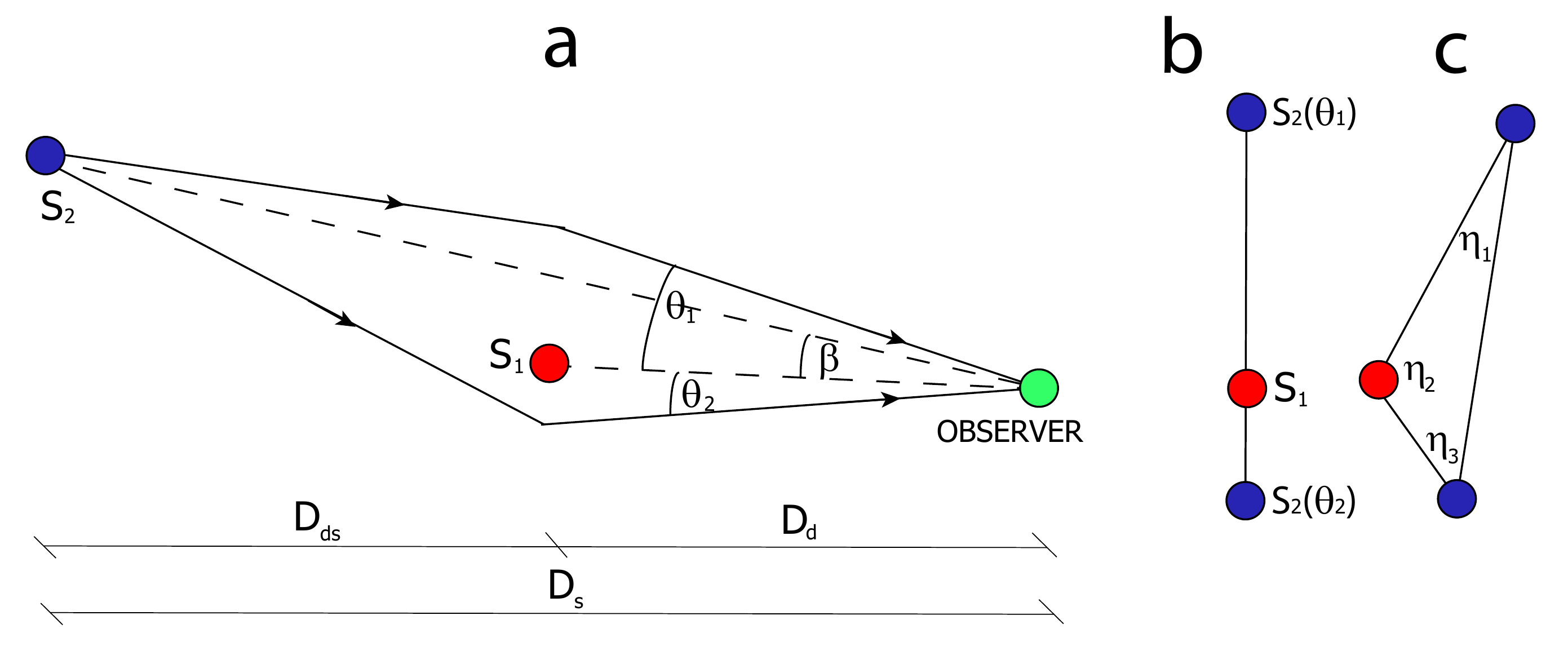}\caption{Geometry of
gravitational microlensing effect (a). Alignment of three sources seen by
observer: perfect (b) and partial (c).}%
\label{faa13}%
\end{figure*}

The principle of the gravitational microlensing effect is explained
in \cite{gl}. For a small angular separation $\beta$ between two stars, the
light beams from the more distant one ($\mathbf{S}_{2}$) and passing the
gravitational field of \ the star on the way ($\mathbf{S}_{1}$) can reach the
observer by the two pathways, as illustrated in Fig.\ref{faa13}a.
The angles seen by the observer are%
\begin{equation}
\theta_{12}=\frac{\beta\pm\sqrt{4\alpha_{0}^{2}+\beta^{2}}}{2};\qquad
\alpha_{0}=\sqrt{\frac{4\varkappa M}{c^{2}}\frac{D_{ds}}{D_{d}D_{s}},}
\label{az34}%
\end{equation}
where $\varkappa$ is a gravitational constant, and $M$ is the mass of the star
$\mathbf{S}_{1}$. For $\beta=0$ the Einstein ring with angular radius
$\theta=\alpha_{0}$ is created. \EMPH{The relation in Eq.(\ref{az34}) implies}
\begin{equation}
\left\vert \theta_{2}\right\vert \leq\alpha_{0}\leq\theta_{1}. \label{az35c}%
\end{equation}
Therefore the necessary condition for observation of the splitting is that
resolving power of the equipment is better than $\alpha_{0}$. For estimation
of $\alpha_{0}$ , first of all the distance $D_{d}$ is critical. For example
$D_{d}\approx10-10^{2}$l.y., $D_{s}\approx2D$ and $M\approx M_{S}$ give
roughly $\alpha_{0}\approx20$ mas. Such a small separation is probably beyond
present Gaia resolution. In fact the minimum separation we have registered
\ in any of the regions C, N, and S is $59$ mas. At the same time, for observation of
the effect it is important that angular separation $\beta$ of the pair is
close to $\alpha_{0}$.\ For $\beta<\alpha_{0}$ , the sources $\mathbf{S}%
_{2}(\theta_{2}),\mathbf{S}_{2}(\theta_{1})$ become strongly magnified,
so the source $\mathbf{S}_{1}$ may not be resolved. On the other hand,
brightness of $\mathbf{S}_{2}(\theta_{2})$ falls rapidly for $\beta>\alpha
_{0}$ \cite{gl}.

\begin{figure*}[t]
\centering\includegraphics[width=16cm]{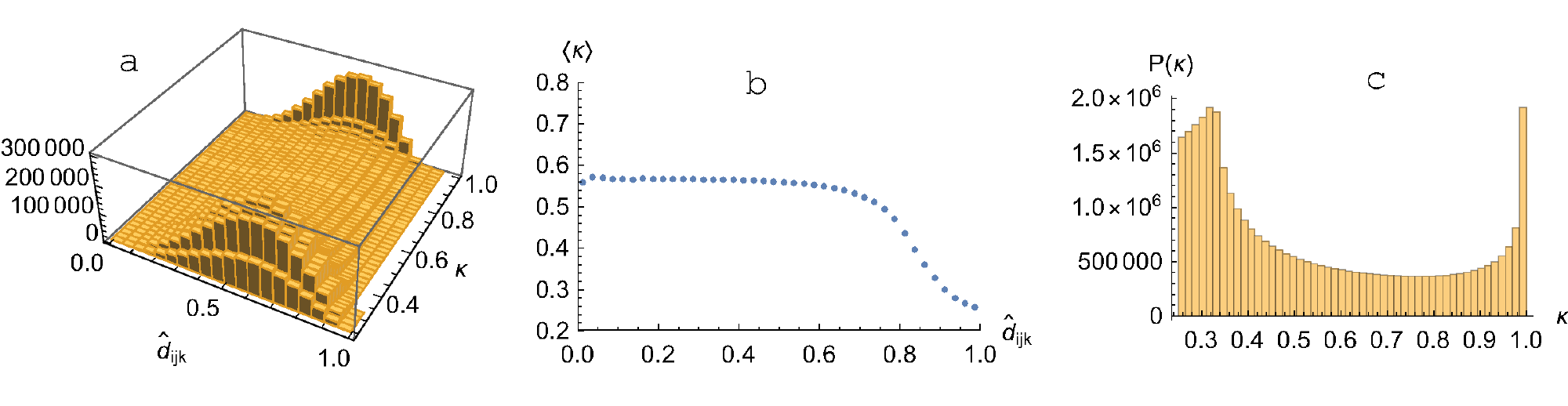}
\includegraphics[height=5cm]{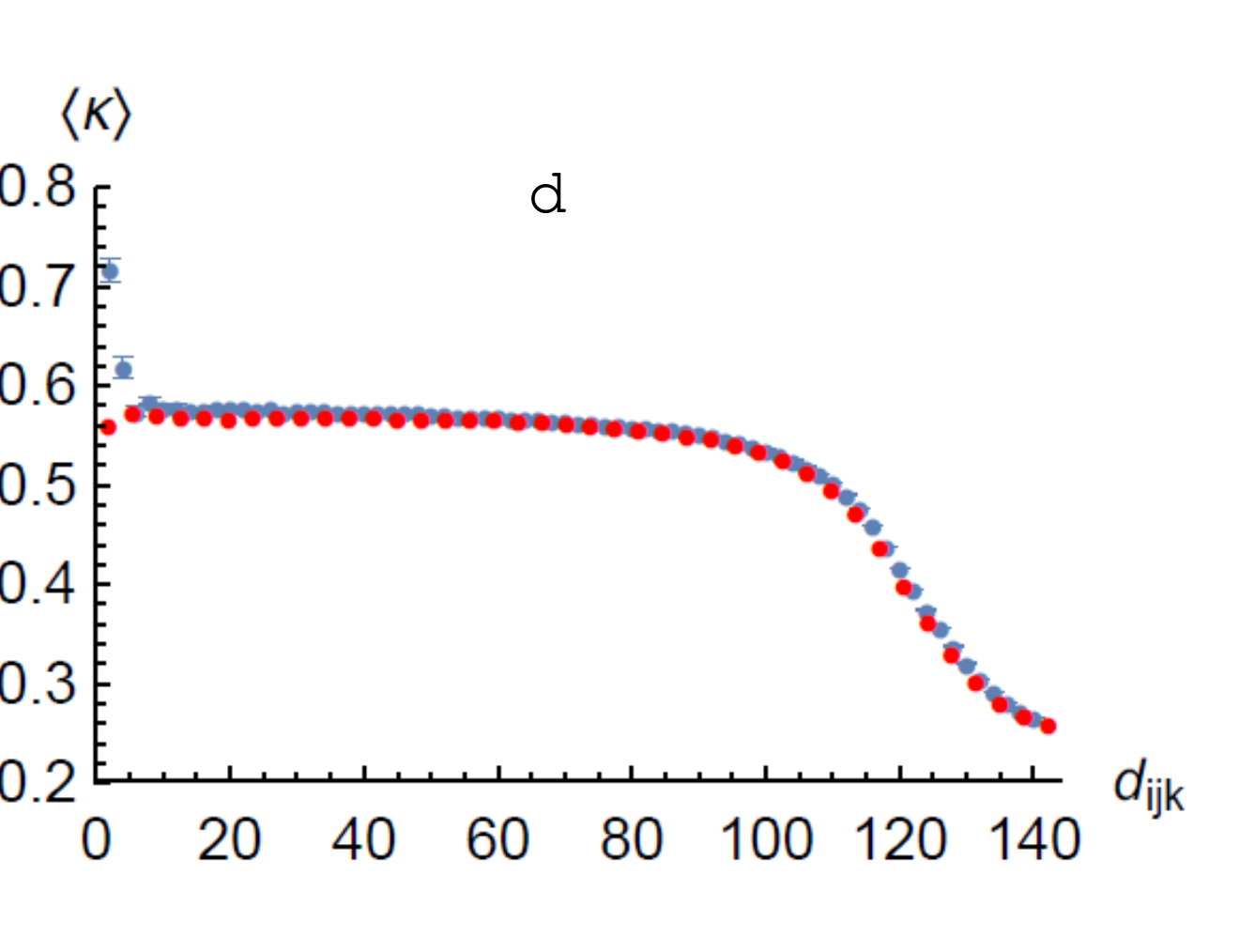}\includegraphics[height=5cm]{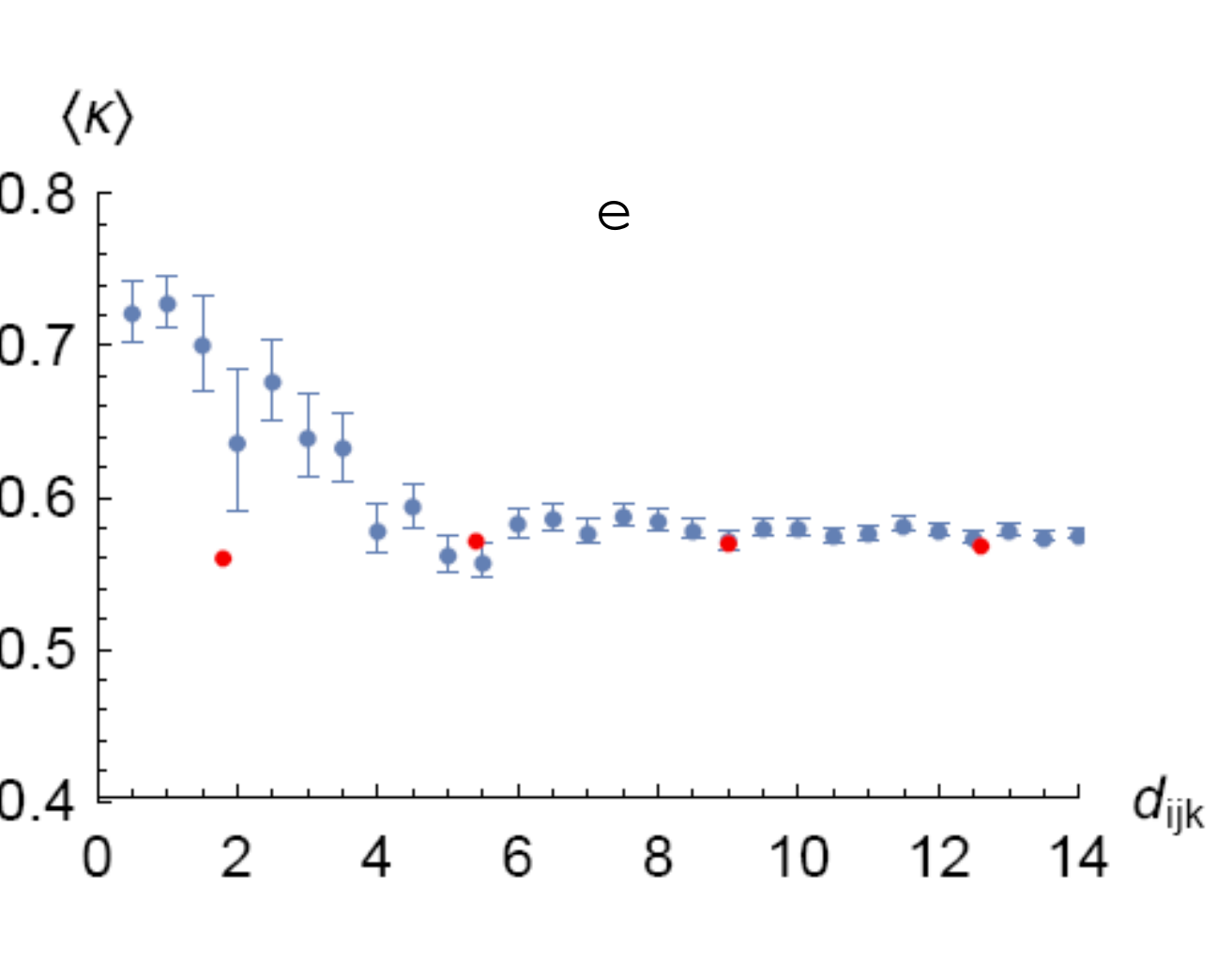}
\caption{The MC simulation of distribution $P(\hat{d}_{ijk},\kappa)$ (a), the
alignment function $\left\langle \kappa\right\rangle $ (b), and integrated
distribution $P(\kappa)$ (c). Panel (d) displays the comparison of
$\left\langle \kappa\right\rangle $ from the data region N\&S (blue) with the
MC simulation (red). Panel (e) shows the same as panel (d) but in higher resolution.}%
\label{faa14}%
\end{figure*}

There can be the following signature of the gravitational microlensing effect.
The light sources $\mathbf{S}_{2}(\theta_{2}),\mathbf{S}_{1},\mathbf{S}%
_{2}(\theta_{1})$ as seen by the observer, should have a small separation
$d_{ijk}$ (or $\hat{d}_{ijk}$) and should be aligned, or make a
narrow triangle within the errors of measurement (Fig.\ref{faa13}b,c). Therefore, as
a measure of the alignment we define the parameter $\kappa$:%
\begin{equation}
\kappa=\frac{1}{3}\left(  \cos^{2}\eta_{1}+\cos^{2}\eta_{2}+\cos^{2}\eta
_{3}\right)  ,\label{az36}%
\end{equation}
where $\eta_{i}$ are angles in the observed triangle. For $\kappa=1$ there is
maximum alignment (e.g., $\eta_{1},\eta_{2},\eta_{3}=0,0,\pi$) and for
$\eta_{1}=\eta_{2}=\eta_{3}=\pi/3$ we have a minimum of $\kappa=1/4$. In the upper
panels of Fig. \ref{faa14} \ we have shown results of the uniform MC
simulation. In the lower part we have shown its comparison with the data from
the \EMPH{region} N\&S (we note the rescaled variable $d_{ijk}=2\rho\hat{d}_{ijk}$).
It is important that the shape of (normalized) MC distribution
$P(\hat{d}_{ijk},$ $\kappa)$ does not depend on the multiplicity of events
$M\geq3$. The reasons are the same as for distributions $P(\hat{d}_{ijk})$\ in
Sect. \ref{dod}. The same holds for the distribution $P(\kappa)$ and the
dependence $\left\langle \kappa\right\rangle $ on $d_{ijk}$. One can observe a
perfect agreement with the data for $d_{ijk}\gtrsim5^{\prime\prime}$. On
contrary, for the smaller distances among the three sources there is a clear
excess of the alignment. Could there be a connection between this effect and the
gravitational microlensing -- image splitting? Or, more probably, is the excess
only another form of the distortion of measurement at small separations?

\section{Summary and conclusion}

\label{sum}We have proposed a general statistical method for the analysis of  finite 2D patterns. Each pattern (event) consists of the stars located inside
the circle of a given radius. We have demonstrated that the method can
identify tiny deviations from uniform distributions; for example, a tendency to
display clustering or anti-clustering.

The method has been applied to the analysis of astrometric data obtained by
the Gaia mission. In parallel with the data, we have generated a large set of
random uniform events using the MC code. In the present study we have focused on
the events of a small radius and correspondingly small multiplicity
($M\lesssim10$). We have shown the functions $\Theta_{n}(\rho,M)$ representing
deviations from random uniform distribution are a useful tool for analysis.
An equally important tool is the set of functions $P\left(  \hat{\xi}\right)  -$
distributions of the parameters characterizing mutual positions and distances
among the stars (doublets and triplets). \ The results of our analysis are
based on the comparison of these functions from both the data and the MC
simulation. The main results are as follows.

1) In the dense \EMPH{region} (C) we observe a 2D representation of reduced resolution
power of two close sources ($d_{ij}\lesssim4^{\prime\prime}$). Resolution
improves for brighter pairs ($G<15$).

2) In the sparse \EMPH{region} (N\&S) we observe an evident excess of the close pairs
($d_{ij}\lesssim9^{\prime\prime}$). The effect is very pronounced for the
bright pairs ($G<15$). A similar effect is observed for the triplets. We
interpret these excesses as the presence of binary and ternary star systems.

3) Apart from these effects we do not observe any violation of the uniformity
on the scale of our events, which is defined by their radius $\rho
=18^{\prime\prime}(72^{\prime\prime})$ for the dense (sparse) \EMPH{region}.

Special attention has been paid to the discussion on the possibility of
detection of the gravitational microlensing and image splitting effect. Our
present conclusion is that the statistical method  suggested here can be a
useful tool for detection of this effect. With the use of the alignment
function $\left\langle \kappa\right\rangle $\ we have observed the excess of
the three-source alignment at separation $<5^{\prime\prime}$, which could
accompany the gravitational image splitting. At the same time we are aware of
some incompleteness in the Gaia survey reported by the Gaia team for this
scale. We believe it will be possible to obtain more consistent results from
the next Gaia data release (DR2).

\vspace{5mm}
\begin{acknowledgements}
This work has made use of data from the European Space Agency (ESA) mission
\textit{Gaia} (\url{https://www.cosmos.esa.int/gaia}), processed by the
\textit{Gaia} Data Processing and Analysis Consortium (DPAC,
\url{https://www.cosmos.esa.int/web/gaia/dpac/consortium}). Funding for the
DPAC has been provided by national institutions, in particular the
institutions participating in the \textit{Gaia} Multilateral Agreement.
The work was supported by the project LTT17018 of the MEYS (Czech Republic).
Further, we are grateful to J.Grygar for deep interest and many valuable
comments, J.Vondrak for critical reading of the manuscript and important comments,
D.Heyrovsky and A.F. Zakharov for illuminating discussions on the gravitational
microlensing effect and O.Teryaev for very useful discussions and inspiring comments.
\end{acknowledgements}

\end{document}